\documentstyle[12pt]{article}

\begin{document}

\begin{center}
{\large\bf CONSTRAINTS ON THE $\pi NN$ COUPLING CONSTANT FROM THE
$NN$ SYSTEM\footnote{Invited talk presented at the {\it Fifth
International Symposium on Meson-Nucleon Physics and the
Structure of the Nucleon}, Boulder (Colorado), Sept.~6-10, 1993.}}
\\
\vspace*{.5cm}
{\sc R. Machleidt
{\rm and} G. Q. Li\footnote{Present address: Cyclotron Institute,
Texas A\&M University, College Station, TX 77843, USA.}}
\\
{\it Department of Physics, University of Idaho,
 Moscow, ID 83843, U.S.A.}
\end{center}

\vspace*{.1cm}
\begin{abstract}
The sensitivity of the deuteron and of $pp$ scattering
to the $\pi NN$ and $\rho NN$ coupling constants is investigated
systematically. We find that the deuteron can be described
about equally well with either {\it large $\pi$ and $\rho$}
or {\it small $\pi$ and $\rho$} coupling constants.
However, $pp$ scattering clearly {\it requires} the strong $\rho$,
but favors the weak $\pi$ (particularly,
in $^3P_0$ at low energies). This apparent
contradiction between bound-state and scattering can be resolved by
either assuming
charge-dependent $\pi NN$ coupling constants
or by adding a heavy pion to the NN model.
In both cases, the neutral-pion coupling constant is small
($g^2_{\pi^0}/4\pi= 13.5$).
\end{abstract}

\vspace*{.5cm}
\noindent
{\bf 1. INTRODUCTION}
\\
Around 1980, it was believed that the $\pi NN$ coupling constant
was well known. Analysing $\pi^\pm p$ data,
Koch and Pietarinen~\cite{KP80}
obtained
$g^2_{\pi^\pm}/4\pi = 14.28 \pm 0.18$
(equivalent to
$f^2_{\pi^\pm}=0.079 \pm 0.001$).
Kroll~\cite{Kro81}
 determined the neutral-pion coupling constant
to be
$g^2_{\pi^0}/4\pi = 14.52 \pm 0.40$
(equivalent to
$f^2_{\pi^0} = 0.080 \pm 0.002$)
from the analysis of $pp$ data
by means of forward dispersion relations.

The picture changed substantially in 1987, when the Nijmegen group~\cite{Ber87}
determined the neutral-pion coupling constant in a partial-wave analysis
of $pp$ data and obtained
$g^2_{\pi^0}/4\pi = 13.1 \pm 0.1$.
Including also the magnetic moment interaction between protons in the analysis,
the value shifted to $13.55 \pm 0.13$ in 1990~\cite{Ber90}.
Triggered by these events, Arndt {\it et al.}~\cite{Arn90} reanalysed
the $\pi^\pm p$ data to determine the charged-pion coupling constant
and obtained
$g^2_{\pi^\pm}/4\pi = 13.31 \pm 0.27$.
These new values for the $\pi NN$ coupling constants do not indicate any
charge-dependence, but they are considerably smaller (by about 6\%)
than the determinations of a decade ago.
In subsequent work, the Nijmegen group analysed also $np$ data~\cite{KSS91}
and $\bar{p} p$ data~\cite{TRS91}. In all cases, they find a ``small''
$\pi NN$ coupling constant.
Their final results are simmarized in Ref.~\cite{STS93}:
$g^2_{\pi^0}/4\pi = 13.47 \pm 0.11$ (equivalent to
$f^2_{\pi^0} = 0.0745 \pm 0.0006$)
and
$g^2_{\pi^\pm}/4\pi = 13.54 \pm 0.05$
(equivalent to
$f^2_{\pi^\pm}=0.0748 \pm 0.0003$).

In all of their work,
the Nijmegen group analyses two-nucleon {\it scattering} data.
The two-nucleon {\it bound state}---the deuteron---is not considered.
On the other hand, the deuteron is very
sensitive to the pion.
Due to the large radius, the long-range interaction provided
by one-pion-exchange (OPE) is dominant in the deuteron~\cite{ER85}.
In particular, the quadrupole moment of the deuteron ($Q_d$)
and the asymptotic $D$-state over $S$-state ratio ($A_D/A_S$)
are direct consequences of the tensor component of the nuclear
force created mainly by the pion.
These deuteron properties are known with an uncertainty
of less than 2\%. Modern realistic models for the NN interaction
(which all use the old ``large'' value for the $\pi NN$ coupling constant,
namely, $g^2_\pi/4\pi\approx 14.4$)
reproduce, in general, the deuteron properties within the empirical
uncertainty of 2\%. Therefore, a substantial reduction of
$g^2_\pi/4\pi$ (by about 6\%) must immediately raise the question
whether we can still explain the deuteron.

\vspace*{.5cm}
\noindent
{\bf 2. THE DEUTERON}
\\
To answer this question, we undertake the following steps.
We start from the Bonn (B) potential~\cite{Mac89}, which we call here
{\bf Model I}.
This model uses for the pion the old ``large''
coupling constant $g^2_\pi/4\pi= 14.4$
and for the $\rho$-meson the large vector-to-tensor ratio
$\kappa_\rho=6.1$~\cite{HP75}. The predictions
by this model
for the crucial deuteron quantities are: quadrupole moment $Q_d=0.278$ fm$^2$
and
asymptotic D-state over S-state ratio $A_D/A_S=0.0264$ (cf.~Table~1).
Note that our predictions for $Q_d$ are based on the
nonrelativistic impulse approximation
and do not include meson-current and relativistic
corrections.  Therefore,
to make the comparison with the experimental data meaningful, we have
subtracted from the experimental value for $Q_d$ [0.2859(3) fm$^2$ \cite{RV75}]
the meson-exchange current and relativistic contributions,
which are 0.010 fm$^2$ for the Bonn potential
according to the most recent and very thorough calculation
by Henning~\cite{Hen93}. Thus, we list 0.276(3) fm$^2$ in the last column
of Table~1 as the empirical quadrupole moment where the assigned error
of 0.003 fm$^2$ is the uncertainty which we assume for the evaluation of the
theoretical corrections. Obviously, Model I (Bonn-B) reproduces the empirical
quadrupole moment quite well.

\begin{table}
\small
{\bf Table 1.} Important coupling constants and the predictions for the
deuteron
and $pp$ scattering for six models considered in this investigation.
The roman numerals in the head of the table refer to the different models
explained in the text.
Note that Model I--III are wrong either in the deuteron or in $pp$ scattering,
while Model IV--VI describe the deuteron {\it and} $pp$ scattering fair
or even well.
\begin{center}
\begin{tabular}{ccccccccc}
\hline\hline
 \hspace*{2.4cm} &~I~&~II~&~III~&\hspace*{.3cm}&~IV~&~V~&~VI
   &\hspace*{.4cm}Experiment\hspace*{.4cm}\\
\hline\hline
\multicolumn{9}{c}{\bf Important coupling constants}\\
{}~$g^2_{\pi^0}/4\pi$~&~14.4~&~13.5~&~13.5~&&~14.0~&~13.5~&~13.5~&~\\
{}~$g^2_{\pi^\pm}/4\pi$~&~14.4~&~13.5~&~13.5~&&~14.0~&~14.4~&~13.5~&~\\
{}~$g^2_{\pi '}/4\pi$~&~0.0~&~0.0~&~0.0~&&~0.0~&~0.0~&~35.0~&~\\
{}~$\kappa_\rho$~&~6.1~&~6.1~&~3.7~&&~6.1~&~6.1~&~6.3~&~\\
\hline
\multicolumn{9}{c} {\bf The deuteron}\\
{}~$Q_d$
(fm$^2$)&~0.278~&~0.266~&~0.274~&&~0.273~&~0.275~&~0.273~&~0.276(3)$^a$~\\
{}~$A_D/A_S$&~0.0264~&~0.0251~&~0.0257~&&~0.0259~&~0.0261~&~0.0256
&~0.0256(4)$^b$~\\
{}~$P_D$ (\%)&~4.99~&~4.56~&~5.31~&&~4.75~&~4.91~&~5.29~&~\\
\hline
\multicolumn{9}{c}{{\bf \mbox{\boldmath $^3P_0$ $pp$} phase shifts} (deg)}\\
10 MeV & 4.014 & 3.731 & 4.105 && 3.883 & 3.731 & 3.727 & 3.729(17)$^c$\\
25 MeV & 9.254 & 8.612 & 9.968 && 8.952 & 8.612 & 8.607 & 8.575(53)$^c$\\
50 MeV &12.39 &11.57 &14.40 && 12.00 &11.57 &11.59 & 11.47(9)$^c$ \\
\hline\hline
\end{tabular}
\end{center}
\footnotesize
$^a)$ Corrected for meson-exchange currents and relativity.
\\
$^b)$ Ref.~\cite{RK90}.
\\
$^c)$ Nijmegen $pp$ multi-energy phase shift analysis~\cite{Sto93}.
\end{table}

In the next step, we lower the pion coupling constant to
$g^2_\pi/4\pi=13.5$, the value suggested by the
Nijmegen analysis~\cite{STS93}: we denote this by {\bf Model II}
(cf.~Table~1). All other parameters that are crucial for
the present discussion are the same as in
Model~I.
(We note that a comparison of two models makes only sense
if the deuteron binding energy
and the triplet effective range parameters are reproduced accurately
and if the over-all fit of the NN phase shifts is satisfactory.
To guarantee this,
the parameters of the $\sigma$-boson and some
short-range parameters are always slightly readjusted in the
development of the various models discussed in this study.)
With this ``small'' pion coupling constant (and $\kappa_\rho=6.1$)
we obtain $Q_d=0.266$ fm$^2$,
which is substantially too low.

We now also lower
the $\rho$-meson coupling to $\kappa_\rho=3.7$ (the so-called
vector-meson dominance model value~\cite{Sak69}), which leads us
to {\bf Model III}. The quadrupole moment now goes up to
0.274 fm$^2$, which agrees well with the empirical value. Thus, by {\it
lowering
both the $\pi$ and $\rho$} coupling constants
we are able to reproduce the deuteron
properties with about the same quality as with
{\it both large $\pi$ and $\rho$} coupling constants.

Thus, no clear-cut decision concerning the pion coupling emerges
from the deuteron (if we assume the $\rho$-coupling to be
uncertain).
We note that we have investigated the deuteron also using other
meson models for the NN interaction and have arrived at the
same conclusions~\cite{MS91}.

\vspace*{.5cm}
\noindent
{\bf 3. PROTON-PROTON SCATTERING}
\\
The natural next step is to turn to NN scattering.
Due to the wealth of partial waves and observables, one
would expect to extract more distinctive
information from scattering.
One question of particular interest is:
Is NN scattering equally indifferent towards the choice
strong-$\pi$+strong-$\rho$ {\it versus} weak-$\pi$+weak-$\rho$
as the deuteron or does it clearly favor one of the two combinations?
If one combination is preferred,
that could yield the decision the deuteron cannot make.
More specifically, one wants to know:
Are there clear indications in NN scattering that
the small pion coupling constant is to be preferred (as found in the
Nijmegen analysis)? And: Does NN scattering clearly favor one of the two
$\rho$ couplings?

Before we address these issues,
we like to raise a slightly different question, namely:
Are there any persistent problems in describing NN scattering
(below about 300 MeV laboratory energy) by meson models?
Many meson-exchange models for the NN interaction have appeared in the
literature over the past 30 years, none of which is perfect (see, e.~g.,
Fig.~5.10 of Ref.~\cite{Mac89} for an overview of how modern
meson-theory based NN models reproduce the NN phase shifts).
In fact, one can identify several typical problems which some meson
models have.
However, most of these problems typically occur at higher energies (200--300
MeV
lab.\ energy) and,
in all of these cases, a solution
of the problem is known (and one can always find a counter example
that does not have the particular
problem). Since the high energy region is sensitive to the short-range part
of the NN interaction,
the solution of these problems comes typically from heavy-boson
exchange, multi-meson exchange, and the parametrization of the
meson-nucleon form factors.

Now, there is one single problem
that occurs at low energies:
essentially all meson models overpredict
the $^3P_0$ phase shifts below 100 MeV, {\it persistently}.
We demonstrate this is Fig.~1(a)  where the solid line represents the
prediction
for the $^3P_0$ $pp$ phase shift by Model~I. We note that the predictions by
other meson models which use the large pion coupling constant are very similar.
Since this is a problem at low energies, heavy-boson
exchange, multi-meson exchange, and form factors cannot cure this problem.
In fact, there is only one single way to solve this problem:
{\it A smaller $\pi NN$ coupling constant}.
The dashed line in Fig.~1(a) shows the result when the small coupling
constant $g^2_\pi/4\pi=13.5$ (Model II) is used;
numerical values for the
$^3P_0$ $pp$ phase shifts for Model~I
($g^2_\pi/4\pi=14.4$) and Model~II ($g^2_\pi/4\pi=13.5$)
are listed in the lower part of Table 1.
It is seen that this small change in the pion coupling constant
has a very large effect on the $^3P_0$ phase shift, and for the smaller
coupling constant (Model~II) there is excellent agreement with the
phase shift analysis.
Other partial waves are little affected by this change of the
$\pi NN$ coupling constant,
and as far as there is a (small) unwanted effect,
it can always be
counterbalanced by very small changes in some of the other
meson parameters.
In summary, the problem of predicting
persistently too large $^3P_0$ phase shifts at low energies
can only be solved by introducing a ``small'' $\pi NN$ coupling
constant, a measure against which other partial waves are
essentially indifferent.

Unfortunately, this does not finish the discussion;
in fact, the real problem starts now.
NN bound state
and NN scattering must, of course, be explained by one and the same
model for the NN interaction.
Now, from our previous discussion we know that the deuteron can tolerate
a small pion coupling constant only if also the $\rho$-meson
coupling is weak.  So when $pp$ scattering needs the weak pion
coupling, then---because of the deuteron---we also have to lower
the $\rho$ coupling, i.~e., we have to use Model III
(weak-$\pi$+weak-$\rho$). The $pp$ phase shifts predicted
by this model are shown by the dotted line in Fig.~1(a)
and Fig.~2 for $^3P_0$ and
$^3P_2$, respectively.
Clearly, when using $\kappa_\rho=3.7$,
the phase shifts of $^3P_0$ and $^3P_2$ come out totally wrong.
Thus, $pp$ scattering requires by all means the strong $\rho$,
but favors the weak $\pi$, a combination
excluded by the deuteron.
We have a problem here.

\vspace*{.5cm}
\noindent
{\bf 4. SATISFYING $pp$ SCATTERING {\it AND} THE DEUTERON}
\\
Presently, we see three possibilities for a solution of the problem:
\begin{enumerate}
\item
One may try a
{\it compromise} between the large and small pion coupling
constant, e.~g., $g^2_\pi/4\pi=14.0$;
we denote this by {\bf Model IV}. The deuteron quadrupole moment then
comes out at the lower limit (cf.~Table~1), while the $^3P_0$ $pp$
phase shifts are already moderately too large [Fig.~1(b)].
It is questionable if one may consider
this model as satisfactory.
\item
One may assume
{\it charge-dependence of the pion coupling constant.} Using
$g^2_0/4\pi =13.5$ for the neutral pion and $g^2_{\pi^\pm}=14.4$
for the charged pion defines our {\bf Model V}.
$^3P_0$ $pp$ is identical and as excellent as in Model II,
but now the deuteron
quadrupole moment is improved to 0.275 fm$^2$,
due to the larger $g_{\pi^\pm}$. Both $pp$ scattering and
the deuteron are described well [cf.~Table~1 and Fig.~1(b)].
\item
Following the suggestion of Ref.~\cite{HT90}, one may include the exchange of
a heavy pion of mass 1200 MeV, $\pi'(1200)$, in the meson model
for the NN interaction, which is done in our {\bf Model VI}.
The additional tensor force provided by $\pi'$ improves the deuteron
just satisfactorily, while it does not deteriorate the $^3P_0$ $pp$
phase shifts.
\end{enumerate}

\vspace*{.5cm}
\noindent
{\bf 5. CONCLUSIONS}
\\
NN scattering requires by all means
a strong $\rho$, consistent with the determination by H\"ohler
and Pietarinen~\cite{HP75}.
There are clear indications in $pp$ scattering to favor a small
neutral-pion coupling constant of $g^2_{\pi^0}/4\pi\approx 13.5$.
The charged-pion coupling constant cannot be pinned down from scattering data
with similar precision, since the $np$ data are typically of lower
quality than the $pp$ data.
However, the deuteron requires a large (charged-)pion coupling
constant unless new tensor-force generating mechanisms
are introduced.

This work was supported by NSF-Grant PHY-9211607 and the Idaho SBOE.

\pagebreak

\begin{center}
{\bf FIGURE CAPTIONS}
\end{center}

\noindent
{\bf Figure 1.} $^3P_0$ phase shifts of proton-proton scattering.
In part {\bf (a)}, the predictions by Model I (solid line), II (dashed),
and III (dotted) are shown; while in part {\bf (b)}, the predictions
by Model IV (solid line), V (dashed), and VI (dotted) are displayed.
The solid dots represent the Nijmegen $pp$ multi-energy phase shift
analysis~\cite{Sto93}.

\vspace*{.5cm}
\noindent
{\bf Figure 2.} $^3P_2$ phase shifts
of proton-proton scattering. The predictions by Model I (solid line),
II (dashed), and III (dotted) are shown. The solid dots represent
the Nijmegen $pp$ multi-energy phase shift analysis~\cite{Sto93}.

\vspace*{3cm}
\noindent
The figures
are available upon request from
\begin{center}
{\sc machleid@idui1.bitnet}
\end{center}
Please, include your FAX number or mailing address with your request.


\begin{thebibliography}{99}
\bibitem{KP80} R. Koch and E. Pietarinen, Nucl. Phys. {\bf A336}, 331
(1980).
\bibitem{Kro81} P. Kroll, in {\it Phenomenological Analysis of
Nucleon-Nucleon Scattering}, Physics Data Vol.~22-1, H. Behrens and
G. Ebel, eds. (Fachinformationszentrum, Karlsruhe, 1981).
\bibitem{Ber87} J. R. Bergervoet, P. C. van Campen, T. A. Rijken,
and J. J. de Swart, Phys. Rev. Lett. {\bf 59}, 2255 (1987).
\bibitem{Ber90} J. R. Bergervoet, P. C. van Campen, R. A. M. Klomp,
J. L. de Kok, T. A. Rijken, V. G. J. Stoks
and J. J. de Swart, Phys. Rev. C {\bf 41}, 1435 (1990).
\bibitem{Arn90} R. A. Arndt, Z. J. Li, L. D. Roper and R. L. Workman,
Phys. Rev. Lett. {\bf 65}, 157 (1990).
\bibitem{KSS91} R. A. M. Klomp, V. G. J. Stoks, J. J. de Swart, Phys.
Rev. C {\bf 44}, R1258 (1991).
\bibitem{TRS91} R. G. E. Timmermans, T. A. Rijken, and J. J. de Swart,
Phys. Rev. Lett. {\bf 67}, 1074 (1991).
\bibitem{STS93} V. Stoks, R. Timmermans, and J. J. de Swart, Phys. Rev.
C {\bf 47}, 512 (1993).
\bibitem{ER85} T. E. O. Ericson and M. Rosa-Clot, Ann. Rev. Nucl. Part.
Sci. {\bf 35}, 271 (1985).
\bibitem{Mac89} R. Machleidt, Adv. Nucl. Phys. {\bf 19}, 189 (1989).
\bibitem{HP75} G. H\"ohler and E. Pietarinen, Nucl. Phys. {\bf B95},
210 (1975).
\bibitem{RK90} N. L. Rodning and L. D. Knutsen, Phys. Rev. C {\bf 41},
898 (1990).
\bibitem{Sto93} V. G. J. Stoks, R. A. M. Klomp, M. C. M. Rentmeester,
and J. J. de Swart, Phys. Rev. C {\bf 48}, 792 (1993).
\bibitem{RV75} R. V. Reid and M. L. Vaida, Phys. Rev. Lett.
{\bf 34}, 1064 (1975); D. M. Bishop and L. M. Cheung, Phys. Rev. A {\bf 20},
381 (1979); T. E. O. Ericson and M. Rosa-Clot, Nucl. Phys. {\bf A405}, 497
(1983).
\bibitem{Hen93} H. Henning, privat communication.
\bibitem{Sak69} J. J. Sakurai, {\it Currents and Mesons}
(University of Chicago Press, Chicago, 1969).
\bibitem{MS91} R. Machleidt and F. Sammarruca, Phys. Rev. Lett. {\bf 66}, 564
(1991).
\bibitem{HT90} K. Holinde and A. W. Thomas, Phys. Rev. C {\bf 42}, 1195
(1990); J. Haidenbauer, K. Holinde, and A. W. Thomas, Phys. Rev. C
{\bf 45}, 952 (1992).
\end{thebibliography}
\end{document}